\documentclass[aps,preprint,prb,amssymb,24pt,floatfix]{revtex4-1}
\usepackage[a4paper,bindingoffset=0.2in,%
      left=1in,right=1in,top=1in,bottom=1in,%
      footskip=.25in]{geometry}
\usepackage{graphicx}
\usepackage{rotating}
\usepackage{dcolumn}
\usepackage{amsmath}
\usepackage{color}
\usepackage{caption}
\usepackage{floatrow}
\usepackage{placeins}
\usepackage{soul}
\usepackage{natbib}
\setcitestyle{numbers,sort&compress,merge,round}
\usepackage[colorlinks=true, allcolors=blue]{hyperref}
\usepackage{soul}

\definecolor{Blue}{rgb}{0,0,1}

\begin{document}

\renewcommand{\thefootnote}{\fnsymbol{footnote}}

\title{Generalized Rayleigh-Plesset Theory for Cell Size Maintenance in Viruses and Bacteria} 

\affiliation{Dept. of Chemistry, University of Texas, Austin, TX 78712}

\author{Abdul N. Malmi-Kakkada$^{1}$, D. Thirumalai$^{1}$
\footnote{Corresponding author: dave.thirumalai@gmail.com}}
\affiliation{$^1$Department of Chemistry, The University of Texas, Austin, TX 78712\\
}
\vspace{2cm}
\date{\today}


\maketitle

\baselineskip24pt



\section*{Abstract}

The envelopes covering bacterial cytoplasm possess remarkable elastic 
properties. They are rigid enough to resist large pressures while being  
flexible enough to adapt to growth under environmental constraints. Similarly, the virus shells play an important role
in their functions. However, the effects of mechanical properties of the outer shell in 
controlling and maintaining the sizes of bacteria or viruses are unknown. 
Here, we present a hydrodynamic ``bubbles with shell" model, motivated by the study of bubble stability in fluids, to demonstrate
that shell rigidity and 
turgor pressure control the sizes of bacteria and viruses. 
A dimensionless compliance parameter, expressed in terms of the elastic modulus of the shell, 
its thickness and the turgor pressure, determines membrane 
response to deformation and the size of the organisms. 
By analyzing the experiment data, we show that bacterial and viral 
sizes correlate with  shell elasticity, which plays a
critical role in regulating size.

\newpage

\section{Introduction}
Viruses consist of genetic material surrounded by a protective coat of proteins called capsids, which   
withstand high osmotic pressures and 
undergo modification (or maturation) to strengthen capsids after viral assembly~\cite{roos2010physical}.
The protective coat is critical in enabling the virus 
to maintain its functionally intact state. 
In bacteria, the envelope covering the cytoplasm, besides being 
essential in sustaining the shape of the cell, 
protects the bacteria from adversary factors such as 
osmotic shock and mechanical stress~\cite{tuson2012measuring,rojas2017homeostatic,Willis17NatRevMicrobiol}. 
Bacterial cell wall is composed mostly of peptidoglycan, whose synthesis, regulation and remodeling
are central to bacterial physiology~\cite{vollmer2008peptidoglycan}. 
Growing body of evidence suggests that proteins controlling the organization of 
peptidoglycan growth could be crucial in the maintenance of cell size~\cite{tuson2012measuring,ouzounov2016mreb}. 
Bacteria and viruses exhibit remarkable diversity in size and shape. 
Nevertheless, for the purposes of developing a physical model, 
we picture them as spherical envelopes enclosing the material necessary for sustaining their lives. 

Individual strains of bacteria are known to maintain  
a narrow distribution of size even when they divide multiple times~\cite{cooper2012bacterial,amir2014cell,ho2015simultaneous}. A number of physical models  
exploring how microorganisms maintain size and shape have been proposed~\cite{thompson1961growth,koch1981surface,boudaoud2003growth,banerjee2016shape}, with similarities between 
cell elongation and bubble dynamics~\cite{thompson1961growth,koch1981surface}. 
Historically, cell size maintenance has been discussed in terms of two major models: ``timer," where 
cells grow for a fixed amount of time before division, and ``sizer,'' where cells commit to 
division at a critical size~\cite{jun2015cell}. Another important model 
is the ``adder'' mechanism, which proposes that a constant size is added between birth and 
division~\cite{amir2014cell,sompayrac1973autorepressor,taheri2015cell}. 
These models incorporate a `license to divide' approach~\cite{campos2014constant} - 
depending upon the passage of time, growth to a specific size or addition 
of fixed size to trigger cell division and regulate size. 
Recently, the need to attain a steady state 
surface area to volume ratio was proposed as the driving factor behind 
size homeostasis in bacteria~\cite{harris2016relative}. 
In emphasizing the need to move away from a `birth-centric' picture, 
alternate models 
relating volume growth to DNA replication initiation have been proposed~\cite{ho2015simultaneous,amir2017point}
based on experiments~\cite{cooper1968chromosome,wallden2016synchronization}. 
Despite significant advances in understanding size homeostasis, the influence of 
important physical parameters of the cell such as  
the turgor pressure and elastic properties of their outer envelope 
on size maintenance is not well known. Even though the molecules that 
control cell cycle and division have been identified~\cite{marshall2012determines},
the ability to predict size from first principles remains a challenging problem.

Here, we develop an entirely different approach by casting the mechanism of size maintenance as an instability problem 
in hydrodynamics. 
We begin by studying the deformation response modes of the cell wall 
using a generalization of the Rayleigh-Plesset (RP) equation, which was derived in the context of modeling the dynamics of 
bubbles in fluids~\cite{brennen2005fundamentals}. The RP equation is a special case of the Navier-Stokes equation used to describe 
the size of a spherical bubble whose radius is $R$. 
We use the term ``shell" generically, being equally applicable to 
membranes, and the composite layers making up the bacterial envelope or capsids. 
In our theory, the shell 
subject to deformation (e.g. expansion) exhibits two fundamental response modes: ($i$)
elastic mode, where perturbative deformation of the cell wall is followed by  
initial size recovery, and ($ii$) unstable response where minute deformation 
results in continuous growth of the deformation. 
The initial size is not recovered in the unstable response mode, and hence 
we refer to it as the plastic response. This is 
similar to the yield point in springs beyond which original length of the 
spring is not recovered after stretching. 
The importance of these two fundamental deformation response 
modes in the context of bending and growth 
in rod shaped cells was investigated recently~\cite{amir2014bending}. 

A key prediction of our theory is the relation between the deformation response modes and  
optimal size, dictated by a single dimensionless compliance parameter, $\zeta$, expressed in terms 
of the elasticity of the shell and the turgor pressure. We show that 
an optimal cell size requires that  $\zeta$  
strike a balance between elastic and plastic response to deformation, thus maintaining 
microorganisms at the edge of stability. 
In general, from a biological perspective, remaining at the edge of stability might facilitate adaptation to changing environmental conditions. 
In fact, this principle is at the heart of `life at the edge' with its consequences found from the molecular~\cite{tartaglia2008prediction} to cellular level~\cite{mora2011biological}.
The model 
consistently predicts the size of sphere-like bacteria and viruses given the physical 
properties of the cell and the protecting shell. 

\section{Theory}
\begin{figure}
\includegraphics[width=0.75\textwidth] {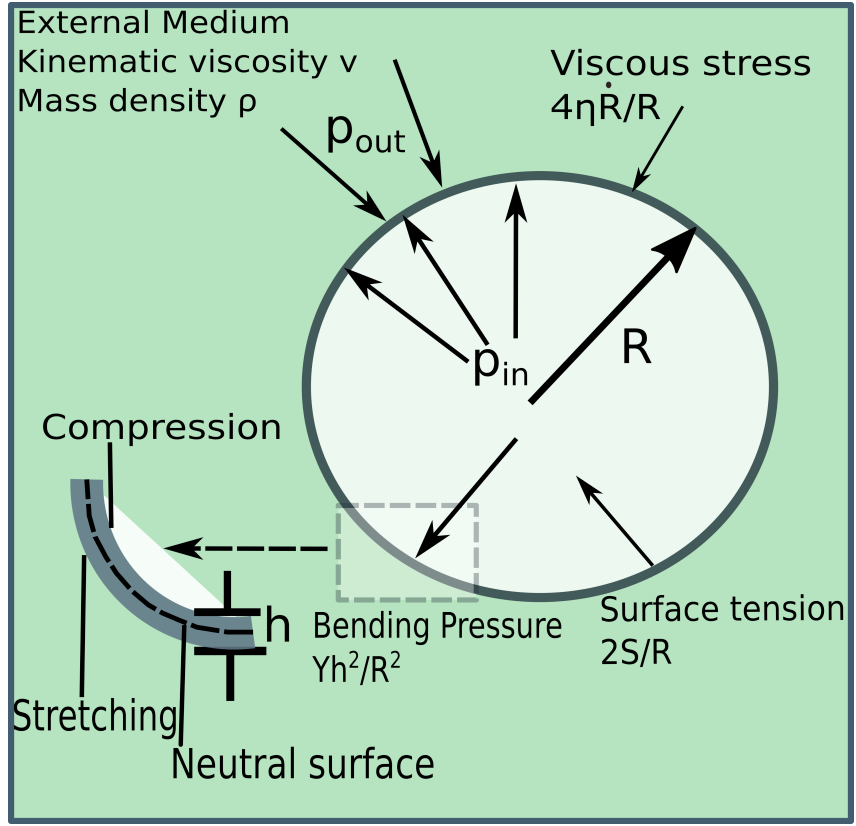} 
\caption{Illustration of the model cell based on the RP equation. The cell wall is a thin shell of 
thickness $h$. The stresses acting on the cell wall are labeled. Viscosity of the surrounding medium is $\eta = \nu \rho$.}
\label{bubbfig}
\end{figure}

Approximating bacteria and viruses as bubbles with shells 
enables us to approach the problem of size maintenance using a  
generalized Rayleigh-Plesset (RP) equation (see Fig.~\ref{bubbfig}). 
For a spherical bubble of radius, $R(t)$, in a liquid at time $t$,  
the temperature and pressure outside the bubble, $T_{out}$ 
and $p_{out}$, are assumed to be constant. The liquid mass density, $\rho$ 
and the kinematic viscosity, $\nu$, are also taken to be constant and uniform. 
If we assume that the contents of the bubble 
are at a constant temperature and exert steady osmotic pressure 
on the bubble wall, $p_{in}$, then the effect of the turgor pressure may be taken into account. 
We outline in Appendix~\ref{motiv} that the RP equation
is motivated from the general equations governing
fluid flow - the continuity and the Navier-Stokes equations.  
To extend the RP equation to study the size maintenance mechanism in microorganisms, 
an additional term for the bending pressure of the thin outer shell is required. 
The elastic energy (per unit area) of bending a thin shell is proportional to the square of the curvature~\cite{deserno2007fluid}.  
Thus, the generalized RP equation is, 
\begin{equation}
\frac{p_{in}(t) - p_{out}(t)}{\rho} +  \frac{Yh^2}{\rho R^2} = R\frac{d^{2}R}{dt^2} + \frac{3}{2}(\frac{dR}{dt})^2 + \frac{4\nu}{R}\frac{dR}{dt} + \frac{2S}{\rho R}, 
\label{rpm}
\end{equation}
where $Yh^2/R^2$ is the bending pressure of the elastic shell, 
$Y$ is the elastic modulus, $h$ is the thickness of the shell, and 
$S$ is the surface tension acting on the shell. 
The bending pressure, or the resistance to bending, arises due to the outer 
side of a bent material being stretched while the inner side is compressed (see Inset in Fig.~\ref{bubbfig}). 
For more details on the bending pressure term see Appendix~\ref{motiv}.
The first term on the left hand side 
accounts for the pressure difference between inside and outside of the cell
and the other terms involve time derivatives of the radius. 

\begin{figure}
\includegraphics[width=.9\textwidth] {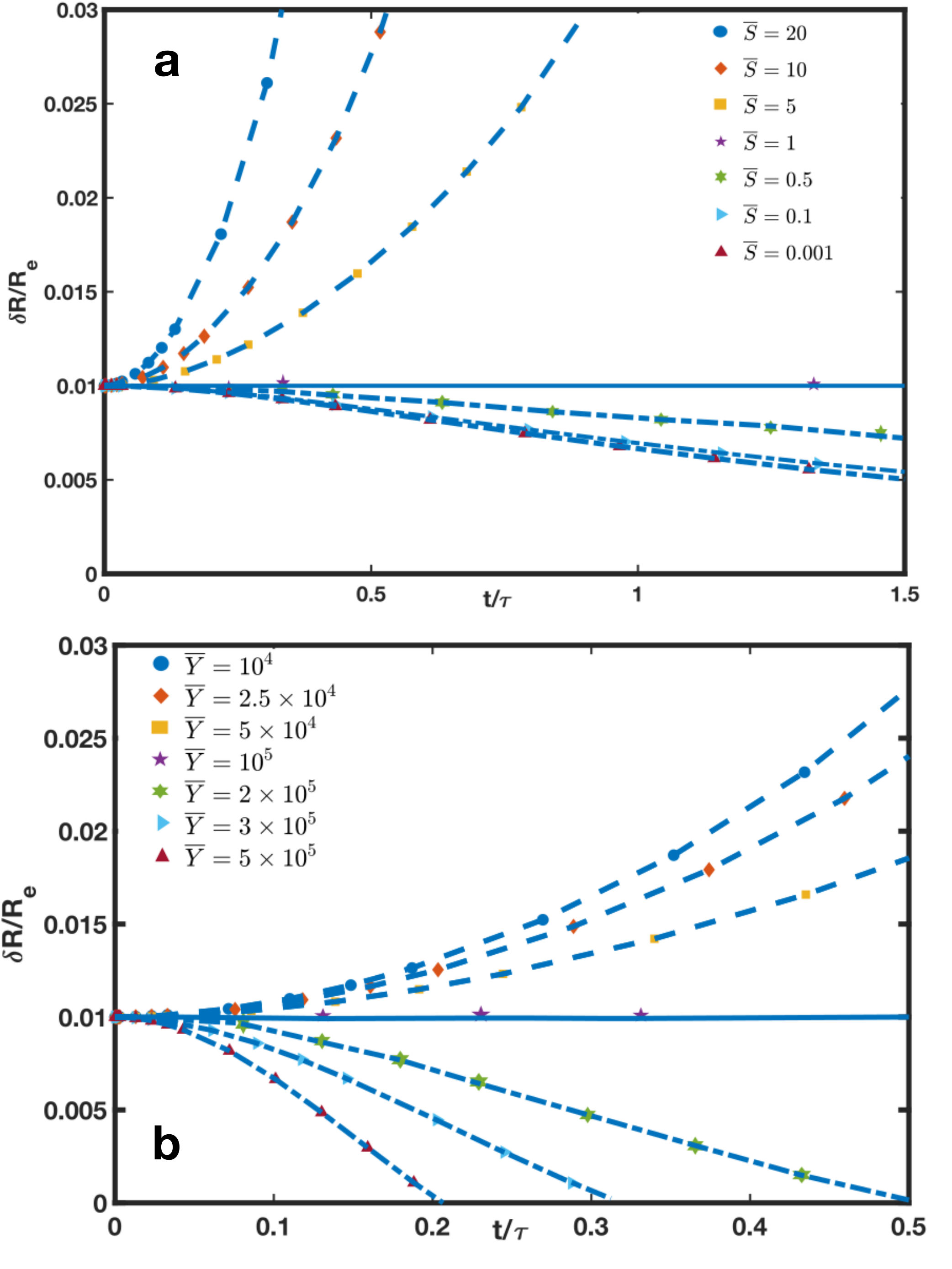}
\caption{Solutions (a) for the first order RP equation shows the time dependent 
behavior of strain. Time is scaled by $\tau$ and length by $R_e$. 
$\bar{Y} = 10^4$ and $\bar{h}=10^{-2}$ are kept constant while $\bar{S}$ is varied.  
(b) Same as in (a) with $\bar{S} =10$ and $\bar{h}=10^{-2}$ kept constant while $\bar{Y}$ is varied.}
\label{fig1}
\end{figure}

Shell displacement, $\delta R(t)$, in the radial direction leads to $R(t) = R_e + \delta R(t)$, 
where $R_e$ is a constant. If $\delta R/ R_e << 1$, an equation for $\delta R(t)/ R_e$ may be derived, 
\begin{equation}
\frac{d^{2}\delta\bar{R}}{d\bar{t}^2} + 4\frac{d\delta\bar{R}}{d\bar{t}} = 2\delta\bar{R}(\bar{S} - \bar{Y}\bar{h}^2), 
\label{rponed}
\end{equation} 
in non-dimensional units where $\delta\bar{R} = \delta R/R_e$, $\bar{h} = h/R_e$ (see Appendix~\ref{ptheory} for further details). 
The stretching energy per unit volume is, $e_{stretch} = \frac{1}{2}Y \int \xi_{\theta}^2 d\Omega$, 
where $d\Omega$ is the differential solid angle, and the angular strain is defined as $\xi_{\theta} = \frac{u_{R}}{R} + \frac{1}{2}(\frac{u_{R}}{R})$~\cite{hannezo2012mechanical}, 
$u_{R}=\delta R$ is the displacement in the radial direction, giving rise to a second order contribution in terms of $\delta R$ which we do not consider here.
We choose $\tau = R_{e}^2/\nu$ which sets the time unit and $R_e$ 
(the mean cell size) is the unit of length. Similarly, the elastic modulus ($\bar{Y}$) and surface tension are rescaled 
using $p_r = \frac{\rho R_{e}^2}{\tau^2}$ and $\bar{S} = S/(p_{r}R_e)$. 
Three types of temporal behavior in $\delta\bar{R}$ are illustrated in Fig.~\ref{fig1}, 
where the radial displacement either increases, stays constant or decays. Both analytic and numerical solutions,  
with initial conditions $\delta\bar{R}(\bar{t}=0)=0.01,~d\delta\bar{R}/d\bar{t} (\bar{t}=0)=0$, may be readily obtained, as detailed in Appendix ~\ref{asoln}. 
Fig.~\ref{fig1}a shows that as the dimensionless surface tension ($\bar{S}$) increases, 
the behavior of $\delta R(t)/ R_e$ changes from continuous decay to growth. 
In Fig.~\ref{fig1}b, $\bar{S}$ is kept constant while the stiffness of the shell is varied.  
Time dependent perturbative displacement, $\delta R(t)/ R_e$, once again 
shows three distinct trends as the shell stiffness, $\bar{Y}$, increases. 
Since $\delta R/ R_e$  is the strain experienced by the elastic shell due to infinitesimal deformation, 
these response modes signify a transition between the `elastic' and the `plastic' regime. 
The plastic regime corresponds to incremental growth in strain while in the elastic regime the strain decays to zero over time, 
implying that the cell size is maintained. The influence of the shell mechanical parameters on oscillation modes are discussed in 
Appendix~\ref{omodes}. 

The strain response modes depend on whether $S$ is greater than or less than $Yh^2/R_{e}$ 
(see Eq.~\ref{rponed}). 
If $S>Yh^2/R_{e}$, continuous growth in strain results in   
the `plastic' regime. However, if $S<Yh^2/R_{e}$,  
a decaying solution for the strain leads to the `elastic' regime. 
The critical value of the surface tension that dictates the boundary 
between the two regimes is predicted to be at $\bar{S}_{c}=\bar{Y}\bar{h}^2$. 
In Fig.~\ref{fig1}a, for $\bar{Y} = 10^4$ and $\bar{h}=10^{-2}$, the critical 
surface tension corresponds to $\bar{S}_{c}=1$. 
Similarly in Fig.~\ref{fig1}b, we show that the critical elastic modulus 
is $\bar{Y}_{c}=\bar{S}/\bar{h}^2=10^{5}$, in agreement with 
numerical results. Note that Table I in Appendix~\ref{edata} shows that the 
parameter ranges considered are physiologically relevant. 
Surface tension forces must be explicitly taken into account 
in studying envelope deformation of bacteria and viruses since the mechanical equilibrium 
of bacterial shells is determined by surface tension~\cite{deng2011direct,arnoldi2000bacterial}. 
Similarly, mechanical properties of viral capsids are determined by surface tension~\cite{zandi2005mechanical} 
or effective surface tension-like terms~\cite{roos2010physical}.

An important prediction of the theory is that the dimensionless compliance parameter,
$\zeta$, quantifies the shell response to perturbative deformation 
and thereby sets a universal length scale for the size of microorganisms.
The parameter $\zeta$ depends on intrinsic physical properties of the cell, 
which collectively play an important role in bacterial and viral shell deformation. 
The details of the derivation of, 
\begin{equation}
\zeta = \frac{Yh^2}{\Delta P R^2},
\label{zeta}
\end{equation}
where  $\Delta P = p_{in} - p_{out}$, are given in the Appendix~\ref{ptheory}. 
The compliance parameter in Eq.~\ref{zeta} may be obtained by equating the bending 
pressure ($\sim~Yh^{2}/R^{2}$ - this form is justified in the Appendix~\ref{motiv}) and the contribution arising from surface 
tension ($\sim~S/R$, the last term in Eq.~\ref{rpm}). By using the Young-Laplace equation 
for $S \sim \Delta P R$, we obtain Eq.~\ref{zeta}. 

Interestingly, the same parameter rewritten as, 
$\kappa=1/\zeta$, was found to be important in distinguishing between 
bending and tension-dominated response of the bacterial wall during the indentation of 
{\it Magnetospirillum gryphiswaldense} with an AFM tip~\cite{arnoldi2000bacterial}. In a more recent 
study~\cite{amir2014bending}, the dimensionless variable $\chi$ (related to the compliance parameter 
by $\chi(R/h)=1/\zeta$), was 
shown to demarcate the boundary between elastic and plastic bending  
regimes for cylindrical bacteria.

The elastic regime corresponds to $\zeta > 1$, while for $\zeta < 1$ the deformation is plastic. 
Since plastic and elastic deformation modes are expected to be of comparable importance in bacterial cell walls~\cite{amir2014bending}, 
we anticipate that the condition $\zeta=1$ could play an important role in determining size. 
Note that $\zeta=1$ corresponds to $\delta R(t)/R_{e} = \mathrm{constant}$ with neither decay nor growth 
in response to perturbative displacement. 
Thus, the boundary between the plastic and elastic regime ($\zeta = 1$) lets us 
identify a critical radius, 
\begin{equation}
R_c^{2} \sim \frac{Yh^2}{\Delta P},
\label{rcrit}
\end{equation}
which is the central result of our work. 

\section{Analysis of Experimental Data}
\begin{figure}
\includegraphics[width=0.82\textwidth] {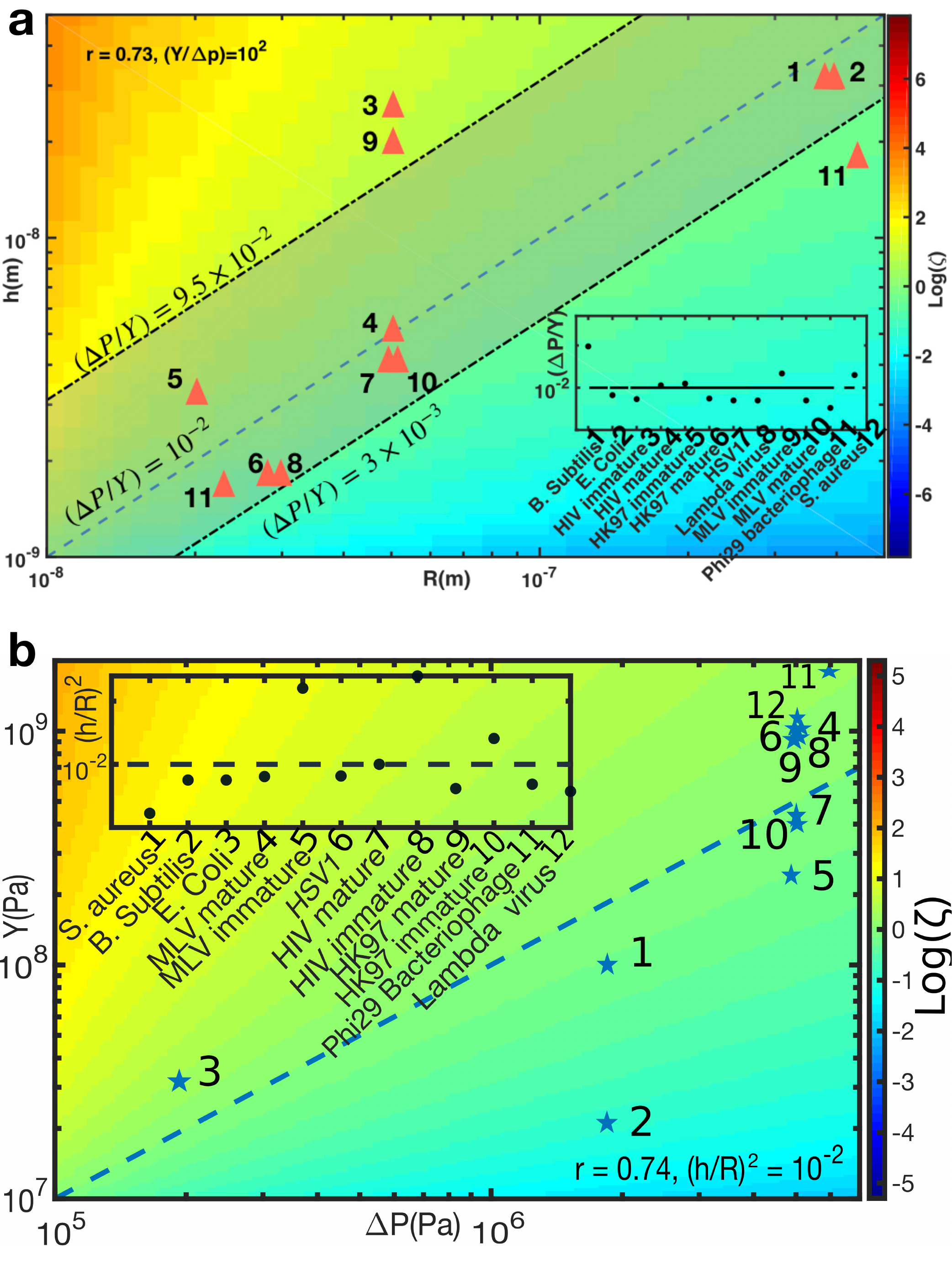}
\caption{(a) Test of the relation given in Eq.~(\ref{thick}) 
for different bacteria and viruses. Inset shows the ratio $\Delta P/Y$. 
(b) Trend in shell elastic modulus ($Y$) versus turgor pressure ($\Delta P$) 
in units of $Pa$. Inset shows the ratio of $(h/R)^{2}$. The data were compiled from existing literature. 
Calculated $\zeta$ (in log scale, with fixed $\Delta P/Y =10^{-2}$ in (a) and $(R/h)^{2} =10^{2}$ in (b)) is indicated in color heat map on the right of the figures. The dashed line in the figure marks $\zeta=1$. }
\label{rpfig}
\end{figure}

The predictions and the ensuing consequences of Eq.~\ref{rcrit} are explored 
by analyzing experimental data. The critical radius obtained above unveils a universal dependence of the size of microorganisms 
on the intrinsic physical parameters of the cell and its outer shell - the pressure difference
between inside and outside, and the elastic modulus and the thickness of the shell, respectively.
We now analyze the size of bacteria ({\it S. aureus, E. coli, B. subtilis}),
and viruses ({\it Murine Leukemia Virus (MLV)},  
{\it $\Phi$29 bacteriophage}, and {\it Human Immunodeficiency Virus (HIV)} etc) in relation to their shell physical properties. 
Data for the radius, elastic modulus, shell thickness 
and pressure difference were obtained from the literature. A comparison of
the shell thickness, $h$, to the radial size, $R$, for 12 bacteria and viruses 
is presented in Fig.~\ref{rpfig}a. 
The thickness of the cell wall is,  
\begin{equation}
h \sim \sqrt{\zeta (\frac{\Delta P}{Y})}R,
\label{thick}
\end{equation}
which is directly proportional to size (based on Eq.~\ref{zeta}). Remarkably, the 
ratio of the turgor pressure to shell stiffness, $\Delta P/Y \sim 10^{-2}$ (see Inset Fig.~\ref{rpfig}a), falls 
on a straight line for most of the bacteria and viruses.  
The ranges of $\Delta P/Y$ on the higher end from $9.5 \times 10^{-2}$ for {\it B. subtilis} and 
on the lower end to $3\times 10^{-3}$, within an order of magnitude. The maximum observed value of 
$\Delta P/Y \sim 9.5\times 10^{-2}$ and the minimum $\Delta P/Y \sim 3\times 10^{-3}$ provides  
an upper and lower bound for our predicted $h$ vs. $R$ relationship, as plotted using dash-dot lines in Fig.~\ref{rpfig}a.
In agreement with our theory, the shell thickness is linearly correlated to the overall size with the Pearson correlation coefficient, $r=0.73$.
As predicted by the theory, the data points lie close to $\zeta =1$ (marked by the dashed line) in Fig.~\ref{rpfig}a,  
indicative of the importance of the balance between plastic and elastic deformation modes in 
microorganisms. 

Because  correlation does not imply causation, we sought to test our prediction using an alternate set of variables.
The relation between turgor pressure and shell elastic modulus 
can be predicted using, 
\begin{equation}
Y \sim \zeta (\frac{R}{h})^{2}\Delta P.
\label{stiff}
\end{equation}
Taking the ratio between radial size and shell thickness, $(R/h)^{2} \sim 10^{2}$ (see Inset Fig.~\ref{rpfig}b), allows us to identify the preferable $\zeta$ regime. 
The Pearson correlation coefficient, $r=0.74$, and the region where $\zeta =1$
is marked by a dashed line in Fig.~\ref{rpfig}b. When either $\Delta P$ or $Y$ were
not available, value of a similar species was used (see Table II in Appendix~\ref{edata} for more details).
Therefore, our conclusion that bacteria and viruses maintain their shell elastic properties 
to lie close to $\zeta =1$ is further borne out by this analysis. 

\begin{figure}
\includegraphics[width=0.8\linewidth] {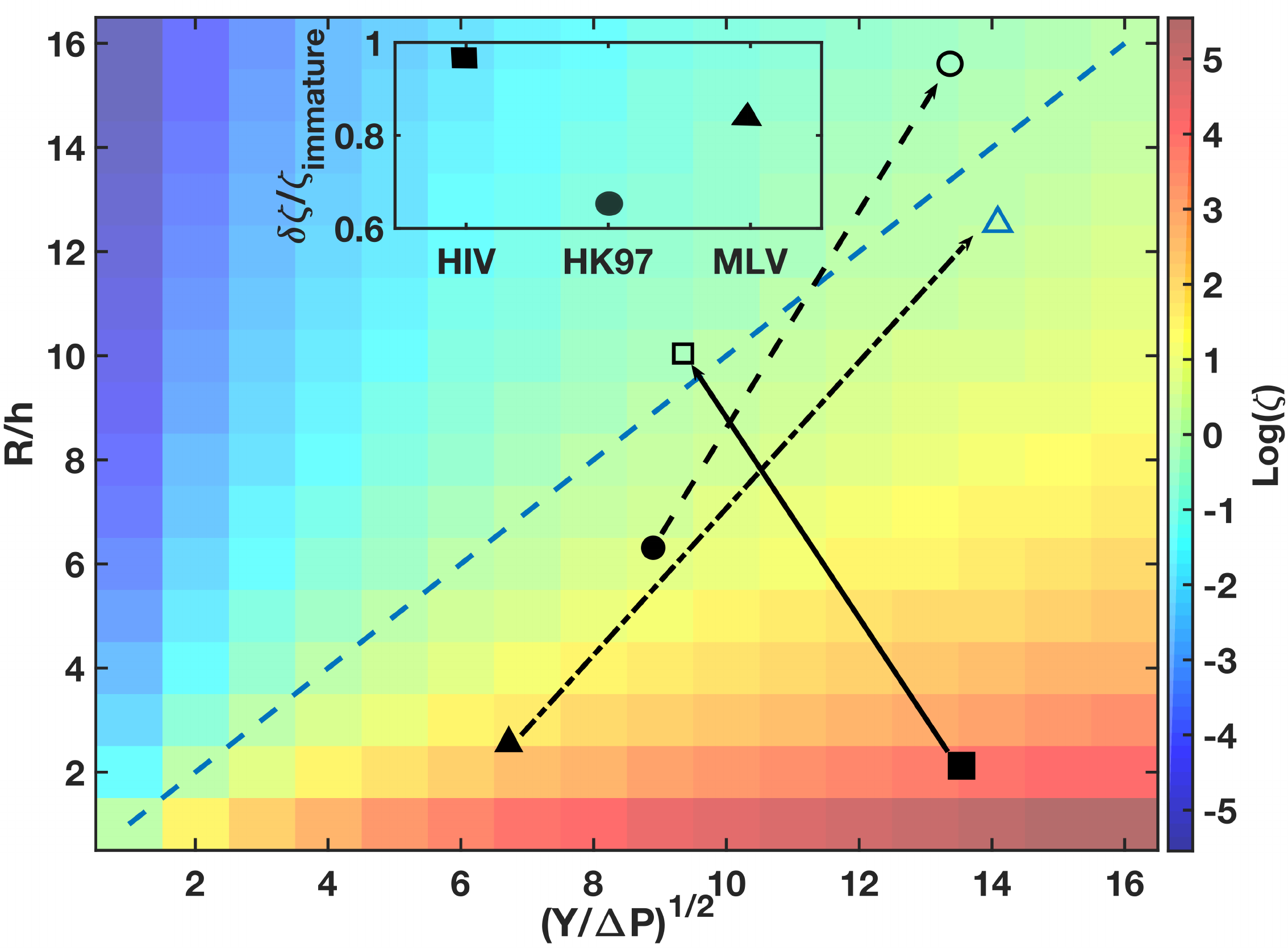} 
\caption{ Comparing the 
ratio of $R/h$ to $(Y/\Delta P)^{1/2}$ during the immature and mature stages of viruses. 
Mature/Immature HIV ($\square$)~\cite{kol2007stiffness},
Mature/Immature HK97 virus ($\bigcirc$)~\cite{roos2012mechanics, johnson2010virus}, 
Mature/Immature Murine Leukemia Virus (MLV $\triangle$)~\cite{kol2006mechanical}. 
Hollow and filled symbols correspond to mature and immature viruses respectively (with lines joining them as guides to the eye). 
Calculated $\zeta$ is indicated in color scale on the right. The dashed diagonal line in the figure marks $\zeta=1$.
Lines connecting filled to hollow symbols visualize the tuning of the physical parameters as a virus matures.}
\label{rpfig1}
\end{figure}

{\it Maturation of Viruses:} We propose that the physical mechanism of adapting to a specific value of $\zeta$ could be utilized by viral particles 
and bacteria to tightly maintain a specific size. We examine the consistency of this proposal by analyzing experimental data. 
We now explore the role of $\zeta$ (Eq.~\ref{zeta}) in the viral maturation process.
Double stranded (ds) DNA bacteriophages are known to undergo 
conformational and chemical changes that tend to strengthen the shell~\cite{roos2010physical} by a 
process that resembles structural phase transition in crystals. This is necessary 
considering that the shells have to be able to withstand large internal pressures and at the same 
time be unstable so that their genome can be released into host cells during infection.
The radius of a viral particle remains approximately constant throughout its life cycle. However, experimental 
evidence shows that shell thickness of viruses is tuned actively (see Table II in Appendix~\ref{edata} and references therein). 
In {\it HIV, MLV} and {\it HK97} viruses, 
the shell thickness decreases during the maturation process~\cite{kol2007stiffness,kol2006mechanical,roos2012mechanics}. Interestingly, 
in {\it MLV} and {\it HK97} the decrease in shell thickness corresponds to an increase in capsid stiffness~\cite{kol2006mechanical,roos2012mechanics}. 
Given these clues, we  quantify the role of $\zeta$ in the viral maturation process.  
Fig.~\ref{rpfig1} compares the size and shell elastic properties of 
individual viruses between their immature and mature phases. 
The tuning of the ratio of radius to the shell thickness to larger values as the virus matures is clearly observed
(filled $\rightarrow$ hollow shapes).
As the viral particle matures, a transition
from the elastic regime($\zeta > 1$) to a plastic regime($\zeta \le 1$) is observed. 
The inset in Fig.~\ref{rpfig1} shows the fractional 
change in the compliance, $\delta \zeta/\zeta = (\zeta_{immature} - \zeta_{mature})/\zeta_{immature}$. 
Notable change in $\zeta$ due to maturation occurs for all three viruses with $\zeta_{mature} << \zeta_{immature}$. 
This marks a crossover behavior in the viral lifecycle where 
surface tension-like forces in the shell begin to dominate the force associated with the shell stiffness. 
As before, the dashed line in the figure ($\zeta = 1$)
separates the elastic regime from the plastic regime. We surmise that such an adaptation is necessary for the onset of 
instability in viral capsids.

The elastic modulus ($Y$) of the viral shell is an important parameter in the maturation process of viruses. 
Shells of {\it MLV} and {\it HK97} become stiffer as they mature. As a result of the increased shell stiffness, 
$h$ decreases while maintaining approximately the same size. 
In the three different viruses analyzed, $\zeta$ approaches $1$,   
which we propose to be a general property of viral maturation. 

{\it Bacteria:} We explore further the consequence of adapting to a specific value of $\zeta$ in a bacterium.
The thickness of the {\it S. aureus} cell wall is tuned to higher values as a result of nutrient depletion in the stationary phase
(in {\it S. aureus} synthetic medium)~\cite{zhou2012nutrient}. Glycine depletion in the nutrient medium
forces {\it S. aureus} to make ``imperfect'' peptidoglycan resulting in a less rigid cell 
wall, which is more susceptible to lysis~\cite{hofkin2010living}. 
{\it S. aureus} responds by increasing the peptidoglycan 
thickness. 
The observed adaptation behavior of the cell wall thickness is to be expected from Eq.~(\ref{rcrit}). 
Given a decrease in $Y$ (due to a defective cell wall) and assuming that the ratio of pressure difference to $Y$ 
is a constant, the bacteria can maintain its size by increasing the thickness, $h$. 
We note that in the stationary phase bacteria arrest their growth and enter dormancy~\cite{gefen2014direct},
maintaining a constant size $R$.

A scaling behavior obtained earlier by balancing the bending pressure of shells with turgor pressure~\cite{boudaoud2003growth} showed that  
$R/h$ is proportional to $(Y/\Delta P)^{1/3}$, focusing on alga cells and fungi. 
This proposed scaling arises due to surface tension-like forces not being considered, 
which perhaps accounts for the departure from 
the proposed scaling for bacteria and viruses (the focus of this study). 
We quantitatively compare the best fit to experimental data presented in Fig.~\ref{rpfig}a 
to the two different scaling behaviors: (i) $h=~\sqrt{\zeta (\Delta P/Y)}~R$, and 
(ii) $h=~\alpha^{-1}(\Delta P/Y)^{1/3}~R$~\cite{boudaoud2003growth} and show 
that the experimental data is better accounted by the generalized RP theory (see Appendix~\ref{compmodel}). 

\section{Conclusion}
By generalizing a hydrodynamic model based on the Rayeigh-Plesset equation, 
originally formulated in the context of bubbles in fluids, we proposed a novel unified framework to predict  
the size of bacteria and viruses from first principles. Given the shell elastic properties 
and the pressure differential between the inside and outside, the importance of selecting a deformation response 
mode is shown as a possible mechanism to constrain size. Nanoscale vibrations, proposed as a signature of life~\cite{kasas2014detecting}, 
could provide a natural basis for bacteria and viruses to detect the elasticity of shells. 
We identified a compliance parameter, $\zeta$, in terms of the physical properties of the cell as the most relevant variable controlling cell size. 
Viral particles are especially sensitive to $\zeta$, and we predict that shell 
properties evolve to minimize $\zeta$ during maturation. 
By merging approaches 
from hydrodynamics and elasticity theory, we have proposed a new mechanism for an important 
question in cell biology pertaining to 
size regulation.
In conjunction with studies on the role of biochemical processes 
in shape and size maintenance, the importance of physical parameters  
should also be considered in order to fully understand size homeostasis in bacteria and viruses. 

\section{Appendix}
\appendix
\section{Motivation of Rayleigh-Plesset Equation}
\label{motiv}

We begin with the equations governing fluid flow in the presence of a bubble (Fig.~1 in the Main Text), the continuity
equation, 
\begin{equation}
\frac{\partial \rho}{\partial t} + \nabla.(\rho \vec{u}) = 0,
\label{cont}
\end{equation}
and the Navier-Stokes equation, 
\begin{equation}
\frac{\partial \vec{u}}{\partial t} + 
(\vec{u}.\nabla)\vec{u} = -\frac{\nabla p}{\rho} + \nu \nabla^2 \vec{u},
\label{ns}
\end{equation}
where $\vec{u}$ is the velocity field. Other parameters are defined in the main text. 
Considering the incompressible fluid limit, the continuity equation requires that, 
\begin{equation}
\vec{u}(r,t) = \frac{F(t)}{r^2} \hat{r}
\label{cont1}
\end{equation} 
where $F(t)$ can be related to $R(t)$
by a boundary condition at the bubble surface. 
Under the assumption that there is no mass transport across the interface of the bubble, 
$u(R,t) = dR/dt $, and we obtain
$F(t) = R^2 \frac{dR}{dt}$.
Since we are interested only in the radial direction, for simplicity, 
we drop the vector sign henceforth. 
Substituting $u$ from Eq.~(\ref{cont1}) into the the Navier-Stokes equation above gives, 
\begin{equation}
-\frac{1}{\rho}\frac{\partial p}{\partial r} = \frac{1}{r^2} \frac{dF}{dt} - \frac{2F^2}{r^5}.
\label{ns2}
\end{equation}
To proceed further, we consider a small, thin segment 
of the bubble liquid interface and 
the net forces (per unit area) acting along the radial direction. They are:
\begin{equation}
\sigma_{rr}|_{r=R} + p_{in} - \frac{2S}{R}= 0
\label{force}
\end{equation}
where $\sigma_{rr}$ is the stress tensor in the fluid, and $S$ is the surface tension of the bubble film. 
For a Newtonian fluid, $\sigma_{rr} = -p + 2\nu \frac{\partial u}{\partial r}$, 
and we obtain
\begin{equation}
p_{in}(t) - p(t)_{r=R} - \frac{4\nu}{R} \frac{dR}{dt} - \frac{2S}{R} = 0.
\label{force1}
\end{equation}
Integrating Eq.~(\ref{ns2}) gives the pressure at the bubble surface
and following the same steps as in chapter 4 of Ref.~\cite{brennen2005fundamentals},
the generalized Rayleigh-Plesset (RP) equation is,
\begin{equation}
\frac{p_{in}(t) - p_{out}(t)}{\rho} = R\frac{d^{2}R}{dt^2} + \frac{3}{2}(\frac{dR}{dt})^2 + \frac{4\nu}{R}\frac{dR}{dt} + \frac{2S}{\rho R}.
\label{rp}
\end{equation}
This equation, in the absence of surface tension and viscous terms, was first derived by Lord Rayleigh~\cite{rayleigh1917pressure} in 1917,
and was first applied to the study of cavitation bubbles by Plesset~\cite{plesset1949thedynamics}. 
\medskip

In order to generalize Eq.~(\ref{rp}) to investigate
size maintenance mechanism in bacteria, an additional 
term accounting for the bending pressure of the thin outer shell is needed. 
The elastic energy (per unit area) of bending is proportional to the square 
of the curvature of the thin shell~\cite{boudaoud2003growth}. The bending pressure, $p_b$,
of the elastic shell is 
\begin{equation}
p_b = \frac{Yh^2}{R^2}
\end{equation}
where $Y$ is the elastic modulus, and $h$ is the shell thickness.  
Therefore, the generalized RP equation that we shall utilize 
in this study is: 
\begin{equation}
\frac{p_{in}(t) - p_{out}(t)}{\rho} +  \frac{Yh^2}{\rho R^2} = R\frac{d^{2}R}{dt^2} + \frac{3}{2}(\frac{dR}{dt})^2 + \frac{4\nu}{R}\frac{dR}{dt} + \frac{2S}{\rho R}.
\label{rpm}
\end{equation}
Stresses acting on the shell are visualized in Fig.~1 of the Main Text. 
\par \bigskip
{\it Bending pressure term:} Here, we motivate why 
the bending pressure term should scale as $Yh^{2}/R^{2}$. 
Consider a thin sheet of material with sides of length $L$, thickness $h$ and bend it so  
that it develops a radius of curvature, $R$. As illustrated in Fig. 1 of the Main Text, one side of the material 
will be stretched while another side is compressed in order to accommodate the bending. 
Assuming that the bending energy of the 
thin sheet is given by, 
\begin{equation}
E_{bend} = \frac{1}{2}Y\frac{(V-V_{0})^2}{V_{0}},
\label{estr}
\end{equation}  
where $Y$ is the elastic modulus of uniaxial extension or compression, $V$ 
is extended or compressed volume and $V_{0}$ the initial volume. Then, the bending energy 
per unit volume is
\begin{equation}
\begin{aligned}
e_{bend} &= \frac{E_{bend}}{L^{2}h} \\
&= \frac{1}{L^{2}h}\int_{0}^{L} \int_{0}^{L} \int_{-h/2}^{h/2} \frac{1}{2}Y\frac{((1+z/R)dx dy dz - dx dy dz)^{2}}{dx dy dz} \\
&= \frac{1}{L^{2}h} \int_{0}^{L} \int_{0}^{L} \int_{-h/2}^{h/2}  \frac{1}{2}Y (z/R)^{2} dx dy dz \\
&\sim Y \frac{h^{2}}{R^2}.
\end{aligned}
\label{bpres}
\end{equation}  
Here, the extended or compressed volume element is 
\begin{equation}
dV=(1+z/R)dx dy dz,
\end{equation}
for a thin sheet bent along the z-direction with radius of curvature, $R$. Note that one of the integrals is over the thickness of the shell, $h$. 

Moreover, Eq.~(2) of Ref.~\cite{boudaoud2003growth} proposes that 
elastic energy {\bf per unit area} for bending is, 
\begin{equation}
E_{b}\sim Eh^{3}/R^{2},
\end{equation}
where $E=Y$, the elastic modulus. 
To obtain the corresponding bending pressure, with 
units of energy per unit volume,  
\begin{equation}
\label{bpres1}
\begin{aligned}
&\frac{\mathrm{Energy~ of ~bending}}{\mathrm{Area}} \sim E_{b} \sim Eh^{3}/R^{2} \\
&\frac{\mathrm{Energy ~of ~bending}}{\mathrm{Volume}} \sim \frac{E_{b}}{h} \sim Eh^{2}/R^{2}.
\end{aligned}
\end{equation} 
The relevant volume is that of the shell, given by $Area \times h$. 
The key point to note here is that  
the volume contribution comes from the integral over the thickness, $h$, of the shell.

\section{Perturbation Theory}
\label{ptheory}

Having motivated the origin of the generalized RP equation, we now investigate the time dependent behavior of radial displacement. 
We consider the displacement of the spherical shell 
along the radial direction as a perturbation, 
\begin{equation}
R(t) = R_e + \delta R(t),
\label{pert}
\end{equation}
with $R_e$ being a constant and $\delta R/ R_e << 1$. 
By substituting Eq.~(\ref{pert}) into Eq.~(\ref{rpm}) we obtain, 
\begin{equation}
\frac{p_{in}(t) - p_{out}(t)}{\rho} +  \frac{Yh^2}{\rho R_{e}^2} =  \frac{2S}{\rho R_e},
\label{rpzero}
\end{equation}
the generalized RP equation to zeroth order. 
To first order in $\delta R$ the generalized RP equation takes the form:
\begin{equation}
R_e\frac{d^{2}\delta R}{dt^2} +  \frac{4\nu}{R_e}\frac{d \delta R}{dt}  = \frac{\delta R}{R_e}(\frac{2}{\rho R_e})(S - \frac{Yh^2}{R_{e}}).
\label{rpone}
\end{equation}
Note that the right hand side of Eq.~(\ref{rpone}) has the same sign as the perturbation 
($\delta R > 0$) if
\begin{equation}
S - \frac{Yh^2}{R_{e}} > 0.
\label{rpst}
\end{equation}
If the condition above holds, the velocity and acceleration of radial growth have the 
same sign as the perturbation implying that 
any small deviation in the radius will cause $R(t)$ to deviate farther away
from $R_e$. We refer to this as the plastic regime. 
Elastic regime is obtained, to linear order, if 
the opposite condition, $S - \frac{Yh^2}{R_{e}} < 0$ holds. 
Using the zeroth order Eq.~\ref{rpzero}, 
we obtain
\begin{equation}
S = \frac{\Delta P R_e}{2} + \frac{Yh^2}{2R_e}
\end{equation}
where $\Delta P = p_{in} - p_{out}$ is the differential pressure between inside and outside. 
Substituting $S$ into Eq.~\ref{rpst} leads to the crucial compliance scale, 
 \begin{equation}
\zeta =  \frac{Yh^2}{\Delta P R_e^2}
\end{equation} 
such that $\zeta < 1$ corresponds to $S - Yh^2/R_{e} > 0$ resulting in the plastic regime, and $\zeta >1$ leads to the elastic regime. 

\medskip

The critical radius $R_c = Yh^2/S$ that separates
the plastic and elastic regime 
can now be re-written as, 
\begin{equation}
R_c^{2} = \frac{Yh^2}{\Delta P},
\label{rcrit1}
\end{equation}
which is equivalent to $\zeta=1$. 
\medskip



\section{Solutions}
\label{asoln}
\begin{figure}
\includegraphics[width=0.8\linewidth] {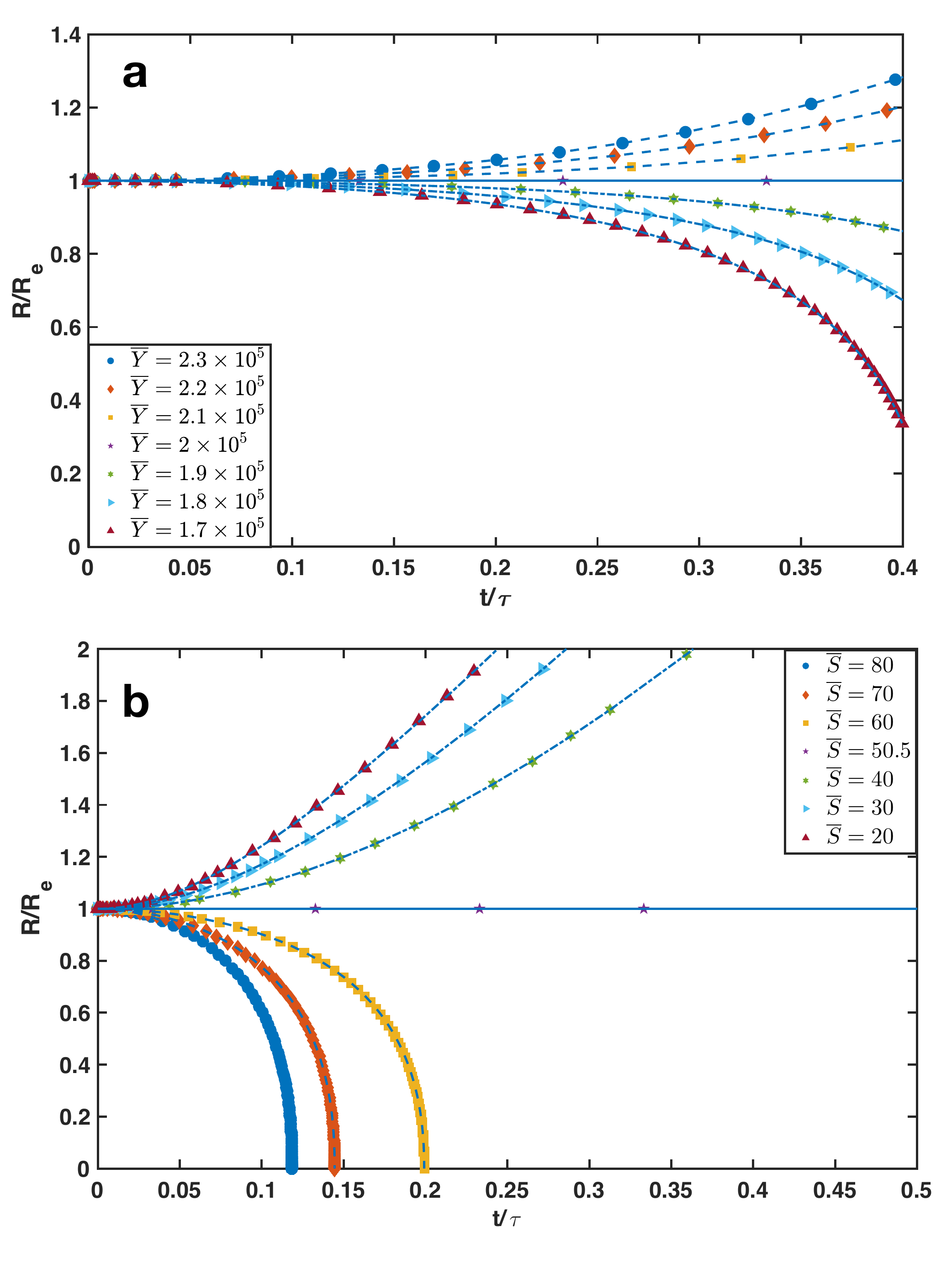} 
\caption{Numerical solutions: (a) Time dependence of the total size $\bar{R}(t)$ from the dimensionless RP equation. Time is scaled by the parameter $\tau$ and length by $R_e$. 
$\bar{S} = 60$ and $\bar{h}=10^{-2}$ are kept constant while $\bar{Y}$ is varied. 
(b) As for (a) but $\bar{Y} =10^4$ and $\bar{h}=10^{-2}$ are constant while $\bar{S}$ is changed. }
\label{figs2}
\end{figure}

Prior to presenting the details of the solution to the RP equation, 
we present a summary of the units. The characteristic 
time scale, $\tau$, is defined in terms of the kinematic viscosity as $\tau = R_{e}^2/\nu$, where $R_e$ rescales the length 
$R$. All pressure terms and surface tension are rescaled 
using $p_r = \rho R_{e}^2/\tau^2$ and $\bar{S} = S/p_r R_e$ respectively. 
Using $\bar{t} = t/\tau$, $\bar{R} = R/R_e$, $\Delta \bar{P} = \Delta P/p_r$, $\bar{Y} = Y/p_r$, 
and $\bar{h} = h/R_e$ one obtains from Eq.~(\ref{rpm}) the dimensionless RP equation, 
\begin{equation}
\bar{R}\frac{d^{2}\bar{R}}{d\bar{t}^2} + \frac{3}{2}(\frac{d\bar{R}}{d\bar{t}})^2 +  \frac{4}{\bar{R}}\frac{d\bar{R}}{d\bar{t}}  = \Delta \bar{P} - \frac{2\bar{S}}{\bar{R}} + \frac{\bar{Y}\bar{h}^2}{\bar{R}^2}. 
\label{rpdim}
\end{equation}
The first order RP equation (Eq.~(\ref{rpone})) in the dimensionless form is,
\begin{equation}
\frac{d^{2}\delta\bar{R}}{d\bar{t}^2} + 4\frac{d\delta\bar{R}}{d\bar{t}} = 2\delta\bar{R}(\bar{S} - \bar{Y}\bar{h}^2).
\label{rponed}
\end{equation} 
In order to solve the equations above, the values of the parameters such as $\tau$, $p_r$, $\bar{S}$ are needed. 
The arbitrary length scale is chosen to be $R_e \sim 10^{-6} \mathrm{m}$ as we focus on the size of microorganisms. 
We use the kinematic viscosity of water as a typical value for  $\nu \sim 10^{-6} \mathrm{m^2/s}$~\cite{asce2013density,venard1975elementary}, and for mass density 
$\rho \sim 10^3 \mathrm{kg/m^3}$. For the surface tension per unit length, we use $S \sim 19 \mathrm{nN/\mu m}$~\cite{banerjee2016shape}.
The magnitude of the quantities $\tau \sim 10^{-6} \mathrm{s}$, $p_r \sim 10^3 \mathrm{N/m^2}$ are thus obtained. 
The initial conditions are $\bar{R}(t = 0) = 1$
and $\frac{dR}{dt}(t=0) = 0$. We use the typical values of $\Delta \bar{P} = 10^2$ and $\bar{h} = 10^{-2}$ for bacteria 
(see Table I for further details). The numerical solution 
(using MATLAB) for $\bar{R}(t)$ from Eq.~\ref{rpdim} above is presented in Fig.~\ref{figs2}. 
The first order Eq.~(\ref{rponed}) is solved both analytically and numerically using MATLAB (ode15s solver). 
The analytical solution to first order RP equation is, 
\begin{equation}
\delta \bar{R}(t) = \frac{(\sqrt{2}+\sqrt{\bar{S}-\bar{Y}\bar{h}^2 +2})}{200(\sqrt{\bar{S}-\bar{Y}\bar{h}^2 +2})}e^{(\sqrt{2(\bar{S}-\bar{Y}\bar{h}^2+2)}-2)t} - \frac{1}{400}(\frac{2\sqrt{2}}{\sqrt{\bar{S}-\bar{Y}\bar{h}^2+2}} - 2) e^{-(\sqrt{2(\bar{S}-\bar{Y}\bar{h}^2+2)}+2)t}. 
\end{equation}
For $\delta \bar{R}(t)$, the radial displacement strain, two behaviors are observed: growth as a function of time 
corresponding to the plastic regime and an elastic regime where $\delta \bar{R}(t)$ decays as a function of time.  A similar behavior is 
also seen for $\bar{R}(t)$. 

\medskip
The bending pressure drives the growth in $\bar{R}(t)$ with larger $\bar{Y}$ leading to continuous increase in size. 
Change in the behavior of $\bar{R}(t)$ from continuous growth to decay either due to a decrease in 
elastic modulus or increase in surface tension of the cell wall is observed. This can be understood from the sign of  
the right hand side of Eq.~\ref{rpdim}, validating the zeroth order Eq.~\ref{rpzero}.

\section{Oscillation Modes}
\label{omodes}
\begin{figure}
\includegraphics[width=0.8\textwidth] {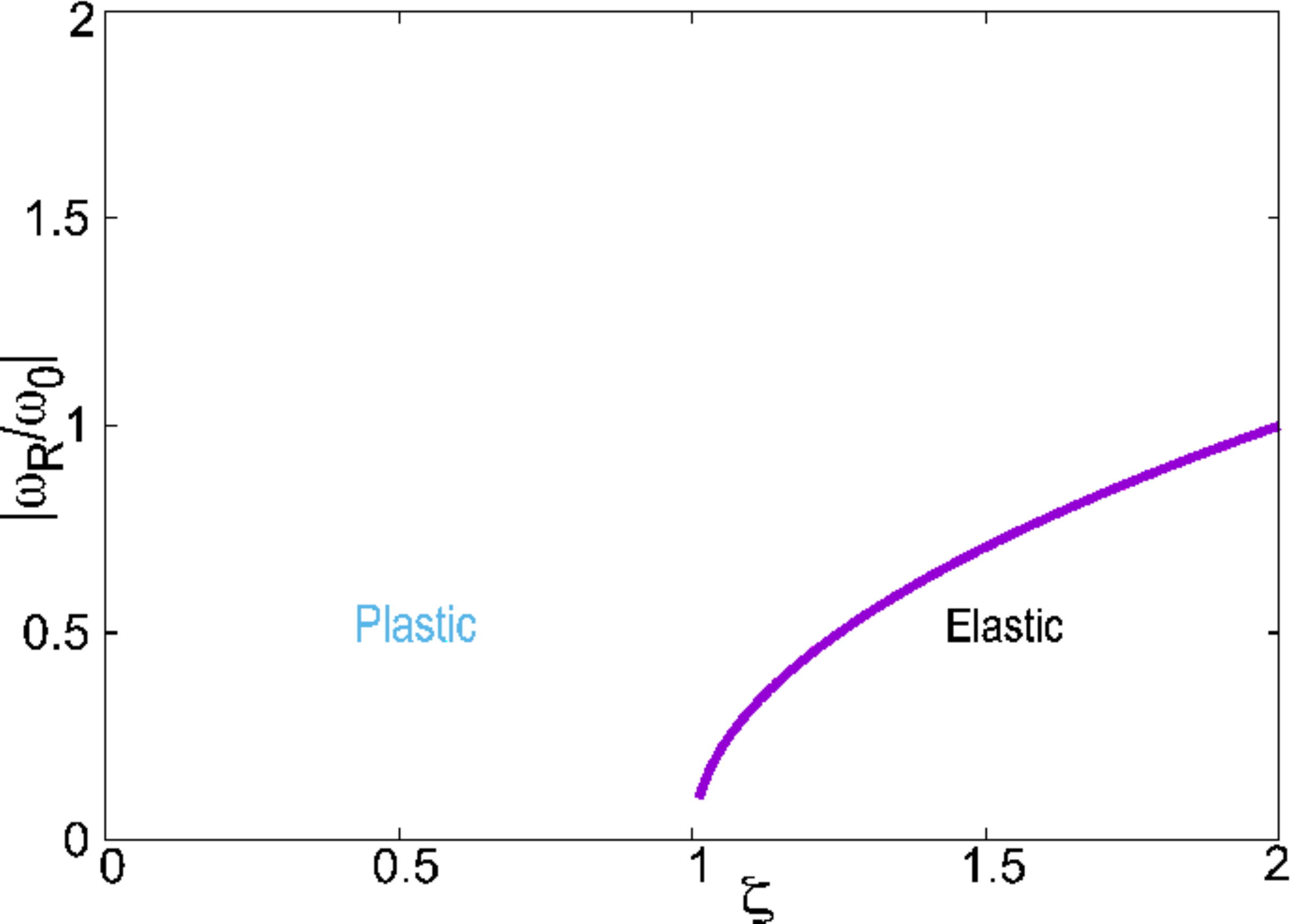} 
\caption{Plot showing the relation between the resonant oscillatory mode and  
$\zeta$. The boundary between the elastic and plastic regimes is indicated. 
Undamped vibrational shell response is not expected to exist in the plastic regime. }
\label{oscfig}
\end{figure}

To study the oscillation modes of the bacterial and viral shell, we 
substitute $\delta R = \delta R_a e^{i\omega t}$ ($\omega$ is the oscillation frequency) into Eq.~(\ref{rpone}) 
and obtain
\begin{equation}
-\omega^2 R_0^2 + 4\nu(i\omega) = (\frac{\Delta P}{\rho} - \frac{Yh^2}{\rho R_{0}^2}).
\end{equation}
Solving for the frequency of the normal modes,
\begin{equation}
\omega = \frac{2i \nu}{R_0^2} \mp \frac{1}{2R_0^2}\sqrt{-16 \nu^2 - 4R_0^2 (\frac{\Delta P}{\rho} - \frac{Yh^2}{\rho R_{0}^2})}.
\label{omega}
\end{equation}
The real part of $\omega$ ($\omega_{R}$) must exist for an oscillatory resonant mode to be present.  
Focusing on the low viscosity limit ($\nu \rightarrow 0$), the resonant oscillation modes 
can only exist in the elastic regime with $\zeta > 1$, which is illustrated in Fig.~\ref{oscfig}. 
Defining $\omega_0 = \sqrt{\Delta P/\rho R_0^2}$,
the dimensionless oscillation frequency of the shell is 
\begin{equation}
\frac{\omega}{\omega_0} = \mp (\zeta - 1)^{1/2}.
\label{oosc}
\end{equation}

 \section{Data}
 \label{edata}
 Numerical values for the various physical parameters considered in Fig~2 of the Main text are summarized in Table I. The values considered 
 for the length and time scale, elastic modulus, surface tension are all shown to be in the relevant physiological range for 
 bacteria. 
  \begin{widetext}
 \begin{table}[]
  \caption{Parameters used to obtain results in Fig~2 of the Main Text.}
  \centering
  \begin{tabular}{c c c c c}
    \hline \hline
    Parameter		&	Value (Dimensionless)		& 		Value (With Dimension)		& 		Experimental Values 	\\ [0.5ex]
    \hline
    $R_{e}$ & --  & $10^{-6}$m & Fixed \\
    $\nu$ & -- & 1.0$\times10^{-6} m^{2}/s$\cite{asce2013density,venard1975elementary}   & --\\
    $\tau(R_{e}^2/\nu)$ & -- & 1.0$\times10^{-6}$sec  & --\\
    $\rho$ & -- &  $10^{3}~kg/m^{3}$\cite{asce2013density,venard1975elementary}   & --  \\
    $p_{r}(\frac{\rho R_{e}^2}{\tau^{2}})$ & --  &  $10^{3}$Pa  & --  \\
    $\bar{Y}(Y/p_{r})$ & $10^{4}~-~5\times10^{5}$  &  $10^{7}~-~5\times10^{8}$Pa   & See Table II  \\
    $\bar{S}(S/p_{r}R_{e})$ & $10^{-1}~-~20$ &  $10^{-4}~-~2\times10^{-2}$N/m   &$10^{-2}~-10^{-1}$N/m \cite{banerjee2016shape,deng2011direct} \\
    $\bar{h}(h/R_{e})$ & 1.0$\times10^{-2}$  &  1.0$\times10^{-8}$   & See Table II\\
    \hline
  \end{tabular}
  \label{table:parameters}
\end{table}
 \end{widetext}
 We summarize the data in Table II for the various physical parameters characterizing bacteria and viruses used to illustrate our theory. 
 \begin{widetext}
 \begin{table}[]
  \caption{Values of parameters used in the model. $*$For thickness of cell wall and membrane.
  Both in Tables I and II the numbers in brackets refer to reference citations.}
  \centering
  \begin{tabular}{c c c c c}
    \hline \hline
    Organism		&		R (m)		& 		h (m) 		& 		$\Delta$P(Pa)		 & 		Y(Pa)  	\\ [0.5ex]
    \hline
    S. aureus & 4.4$\times 10^{-7}$\cite{milo2009bionumbers} & 1.8$\times10^{-8}$\cite{milo2009bionumbers} & 1.9$\times10^6$\cite{milo2009bionumbers,mitchell1956osmotic} & 9.5$\times 10^7$\cite{tuson2012measuring}\\
    B. subtilis & 4.0$\times 10^{-7}$\cite{amir2014bending} & 3.0$\times10^{-8}$\cite{amir2014bending}   & 1.9$\times10^6$\cite{amir2014bending} & 2.0$\times10^7$\cite{amir2014bending}\\
    E. coli & 4.0$\times 10^{-7}$\cite{amir2014bending} & 3.0$\times10^{-8~*}$\cite{hobot1984periplasmic}  & 2.0$\times10^5$\cite{amir2014bending} & 3.0$\times10^7$\cite{amir2014bending}\\
    Murine Leukemia Virus (mature) & 5.0$\times10^{-8}$\cite{kol2006mechanical}  &  4.0$\times10^{-9}$\cite{kol2006mechanical}   & 5.0$\times10^{6}$\cite{zandi2005mechanical}   & 1.0$\times10^{9}$\cite{kol2006mechanical}  \\
    Murine Leukemia Virus (Immature) & 5.0$\times10^{-8}$\cite{kol2006mechanical}  &  2.0$\times10^{-8}$\cite{kol2006mechanical}   & 5.0$\times10^{6}$\cite{zandi2005mechanical}   & 2.3$\times10^{8}$\cite{kol2006mechanical}  \\
    Herpes Simplex Virus 1 & 4.95$\times10^{-8}$\cite{roos2010physical, roos2009scaffold}  &  4.0$\times10^{-9}$\cite{roos2010physical, roos2009scaffold}   & 5.0$\times10^{6}$\cite{zandi2005mechanical}   & 1.0$\times10^{9}$\cite{roos2010physical, roos2009scaffold} \\
    HIV (mature) & 5.0$\times10^{-8}$\cite{kol2007stiffness}  &  5.0$\times10^{-9}$\cite{kol2007stiffness}   & 5.0$\times10^{6}$\cite{zandi2005mechanical}   & 4.4$\times10^{8}$\cite{kol2007stiffness}  \\
    HIV (Immature) & 5.0$\times10^{-8}$\cite{kol2007stiffness}  &  2.5$\times10^{-8}$\cite{kol2007stiffness}   & 5.0$\times10^{6}$\cite{zandi2005mechanical}   & 9.3$\times10^{8}$\cite{kol2007stiffness}  \\
    HK97 (mature) & 2.8$\times10^{-8}$\cite{roos2012mechanics}  &  1.8$\times10^{-9}$\cite{roos2012mechanics}   & 5.0$\times10^{6}$\cite{zandi2005mechanical}   & 9.0$\times10^{8}$\cite{roos2012mechanics}  \\
    HK97 (Immature) & 2.0$\times10^{-8}$\cite{roos2012mechanics}  &  3.2$\times10^{-9}$\cite{roos2012mechanics}   & 5.0$\times10^{6}$\cite{zandi2005mechanical}   & 4.0$\times10^{8}$\cite{roos2012mechanics}  \\
    $\Phi$29 bacteriophage & 2.3$\times10^{-8}$\cite{carrasco2011built}  &  1.6$\times10^{-9}$\cite{carrasco2011built}   & 6.0$\times10^{6}$\cite{carrasco2011built}   & 1.8$\times10^{9}$\cite{carrasco2011built}  \\
    $\lambda$ virus & 2.95$\times10^{-8}$\cite{ivanovska2007internal}  &  1.8$\times10^{-9}$\cite{roos2010physical}   & 5.0$\times10^{6}$\cite{zandi2005mechanical}   & 1.0$\times10^{9}$\cite{ivanovska2007internal}  \\   
    \hline
  \end{tabular}
  \label{table:rp}
\end{table}
 \end{widetext}
 
\section{Comparison to An Existing Model}
\label{compmodel}
\begin{figure}
\includegraphics[width=0.8\linewidth] {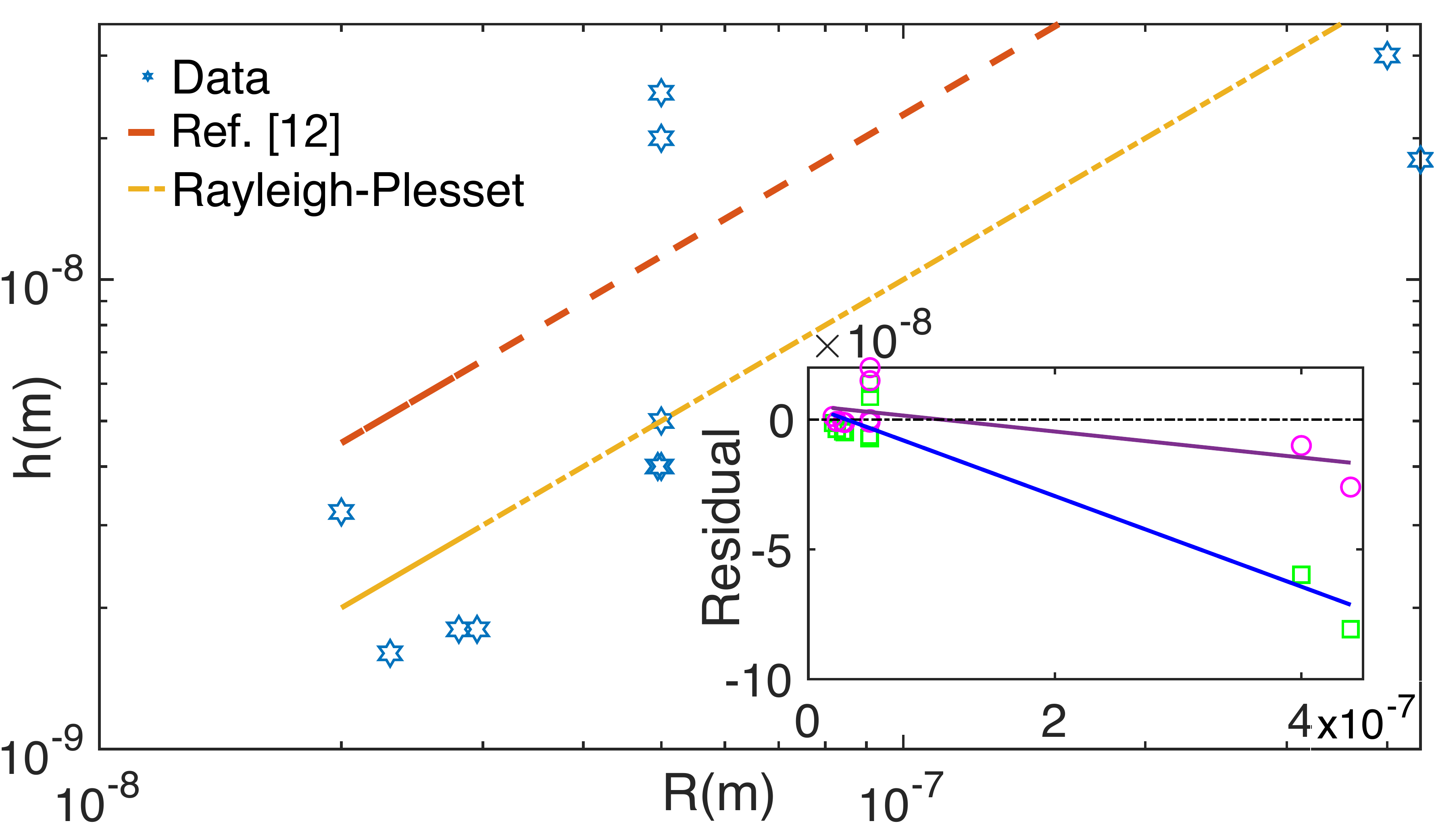} 
\caption{Shell thickness ($h$) for organisms of radial size $R$. 
The red dashed line is the proposed scaling of shell thickness according to the theory 
proposed by Ref.~\cite{boudaoud2003growth}. The orange dashed-dot line is calculated from the 
generalized Rayleigh-Plesset theory. Inset: Residuals from theory in Ref.~\cite{boudaoud2003growth} (square) and 
generalized Rayleigh-Plesset theory fit (circle) as a function 
of $R$. Residuals are fit merely to function as guides for the eye.}
\label{residual}
\end{figure}

By analyzing the data in Fig.~3 (Main Text), we assess whether the 
variability/spread in data allows us to compare our theory 
and the one proposed by Boudaoud~\cite{boudaoud2003growth}. 
Two different scaling laws are put to test: (i) $h=~\sqrt{\zeta (\Delta P/Y)}~R$, and 
(ii) $h=~\alpha^{-1}(\Delta P/Y)^{1/3}~R$ ($\alpha$ being a constant~\cite{boudaoud2003growth}). 
Even though the variation of $h$ with $(\Delta P/Y)$ is very different between the two models, the 
scaling laws relating thickness of the shell, $h$, and cell size, $R$, are of the form, 
\begin{equation}
h=A \times R, 
\end{equation}
where the coefficient $A$ ($A_{B}=\alpha^{-1}(\Delta P/Y)^{1/3}$ or $A_{RP}=\sqrt{\zeta (\Delta P/Y)}$) is to be determined. 
Fitting the data based on the theory proposed in Ref.~\cite{boudaoud2003growth} 
was found to be quantitatively worse (see Fig.~\ref{residual}), based on residuals, $h(R)-\tilde{h}(R)$ and 
percentage errors. Here, $\tilde{h}(R)$ represents the shell thickness values predicted from theory.  

\par \medskip
The coefficient $A$ is a function of the parameter $\Delta P/Y$. 
As shown in the inset of Fig.~3a (Main Text), for the microorganisms 
considered, the ratio of the turgor pressure to shell stiffness is well approximated by $\Delta P/Y~ \sim ~10^{-2}$. 
Therefore, we estimate the coefficient $A$ and compare it to best fit of the experimental data. 
For the experimental data, best fit to shell thickness versus $R$ is obtained with coefficient $A=0.053$ with a $95\%$ 
confidence interval (CI) spanning the lower limit of $0.0198$ and an upper limit of $0.0858$.
\par \bigskip
\begin{tabular}{|p{4cm}||p{4cm}|p{4cm}|p{4cm}|} 
 \hline
\bf{ Fit Type} & \bf{Coefficient A} & \bf{Best Fit(95$\%$ CI)} & \bf{Reference} \\ [0.5ex] 
 \hline
 $h=\sqrt{\zeta (\Delta P/Y)}R$ & 0.1 & $0.053(0.0198~-~0.0858)$ & This paper\\ 
 \hline
 $h=\alpha^{-1}(\Delta P/Y)^{1/3}R$ & 0.37($\alpha=0.58$) to 0.22($\alpha=0.96$) & $0.053(0.0198~-~0.0858)$ & Eq.(4) Ref.\cite{boudaoud2003growth} \\ 
 \hline
 \end{tabular}
\par \medskip
Error analysis reveals that the scaling relation proposed by our generalized Rayleigh-Plesset model 
gives a better agreement with experimental data, which is likely due to the importance of the surface tension. 
Considering the value of $A=0.0858$ at the upper 
limit of the CI, we estimate the error $\Delta$ (expressed as a $\%$) as,  
\begin{equation}
\begin{aligned}
\mathrm{\Delta_{B}}&=(A_{B}-A_{best fit})/A_{best fit}, \\
&=(0.22-0.0858)/0.0858,\\
&=156\%
\end{aligned}
\end{equation}
For the generalized R-P theory, the error is given by, 
\begin{equation}
\begin{aligned}
\mathrm{\Delta_{RP}}&=(A_{RP}-A_{best fit})/A_{best fit}, \\
&=(0.1-0.0858)/0.0858,\\
&=16.5\%
\end{aligned}
\end{equation}
Therefore, quantitatively we conclude that the experimental data 
for the sizes of bacteria and viruses are better accounted for by the generalized 
Rayleigh-Plesset theory proposed here. For microscopic cell sizes of $R < 1\mu m$, Ref.~\cite{boudaoud2003growth} 
notes that the experimental data departs from the theoretical scaling while 
good agreement is observed for cells of size in the range $1-100 \mu m$. Finally, it will be most interesting to perform experiments 
to measure $h$ by changing $\Delta P$ while keeping $Y$ constant or vice versa. This would distinguish between the 
generalized RP prediction and the 1/3 scaling proposed in Ref.~\cite{boudaoud2003growth}.

{\bf Acknowledgements:} This work was supported by the National Science
Foundation through NSF Grant No. PHY 17-08128. 
Additional support was provided by 
the Welch Foundation through the Collie-Welch Chair (F-0019). 
We are grateful to Mauro Mugnai, and Xin Li for discussions and comments on the manuscript.

{\bf Data availability:} The authors declare that the data supporting the findings of this study are available 
within the paper [and its supplementary information files]. 


\clearpage
\bibliography{Cell_Sizerev5} 
\bibliographystyle{unsrt}

\end{document}